# COOLING FLOWS AND METALLICITY GRADIENTS IN CLUSTERS OF GALAXIES

Andreas Reisenegger[1], Jordi Miralda-Escudé[1,2], and Eli Waxman[1]
E-mail: andreas@sns.ias.edu, jordi@sns.ias.edu, waxman@sns.ias.edu

## ABSTRACT

The X-ray emission by hot gas at the centers of clusters of galaxies is commonly modeled assuming the existence of steady-state, inhomogeneous cooling flows. We derive the metallicity profiles of the intracluster medium expected from such models. The inflowing gas is chemically enriched by type Ia supernovae and stellar mass loss in the outer parts of the central galaxy, which may give rise to a substantial metallicity gradient. The amplitude of the expected metallicity enhancement towards the cluster center is proportional to the ratio of the central galaxy luminosity to the mass inflow rate. The metallicity of the hotter phases is expected to be higher than that of the colder, denser phases. The metallicity profile expected for the Centaurus cluster is in good agreement with the metallicity gradient recently inferred from ASCA measurements (Fukazawa et al. 1994). However, current data do not rule out alternative models where cooling is balanced by some heat source. The metallicity gradient does not need to be present in all clusters, depending on the recent merging history of the gas around the central cluster galaxy, and on the ratio of the stellar mass in the central galaxy to the gas mass in the cooling flow.

*Subject headings*: Cooling Flows — Galaxies: Clusters: General — Galaxies: Clusters: Individual (Centaurus, A 3526) — Galaxies: Individual: NGC 4696 — X-Rays: General

---

[1] Institute for Advanced Study, Princeton, NJ 08540
[2] Institute for Theoretical Physics, University of California, Santa Barbara, CA 93106

## 1. INTRODUCTION

X-ray observations in clusters of galaxies have shown that the central gas often has short cooling times (Edge et al. 1992), and that the X-ray spectra show an excess of low-energy photons compared to single temperature spectra. This excess may be understood as arising from cold gas components at temperatures well below the average X-ray temperature. It is usually inferred from these observations that the gas is cooling and that a "cooling flow" forms (see, e.g., Fabian 1994 for a review).

When interpreted in terms of steady-state cooling flow models, the X-ray surface brightness profiles in cooling regions yield a mass inflow rate that increases with radius, requiring an inhomogeneous medium with phases at different temperatures at every radius, generating a distributed mass drop-out as some of the gas cools below X-ray temperatures (Nulsen 1986; Thomas, Fabian, & Nulsen 1987). However, it has not been proved that such a multiphase medium can be maintained against buoyancy forces; moreover, the presence of a gas inflow has not been directly demonstrated, and the fate of the cooled gas remains unknown. A possible alternative is the presence of an energy source that can balance cooling, and for which several alternatives have been proposed (Tucker & Rosner 1982; Silk et al. 1986; Miller 1986; Pringle 1989). This could lead to a cycle where the heated gas could rise buoyantly from the center, while cooled gas could form clouds which would fall and re-evaporate, with no need for a net inflow or mass deposition (see Tabor & Binney 1993).

Recently, a metallicity gradient was observed in the hot gas of the Centaurus cluster (Fukazawa et al. 1994). In this *Letter* we consider the expected metallicity profile in inhomogeneous cooling flow models, as well as in alternative models with no net inflow. Although it might seem that in cooling flow models one should not expect metallicity gradients since the gas is flowing in from large radius, we show that a substantial metallicity gradient may result due to metal enrichment of the inflowing gas by the central galaxy.

## 2. COOLING FLOW MODELS

Inhomogeneous cooling flow models (see Nulsen 1986) assume that at all radii there is a multiphase medium with a distribution of temperatures. Relative motions of gas in different phases due to buoyancy and heat exchange among phases are prevented by a tangled magnetic field. While it has not been shown that these assumptions can be valid in reality, they are the only known way to reconcile steady-state cooling flow models without energy injection with the observed X-ray profiles and spectra. The phases that cool to low temperatures are "deposited" and no longer contribute to the average weight of the multiphase medium, so the mass inflow rate $\dot{M}(r)$ increases with radius.

Cooling flows generally occur around central cluster galaxies, which often have extended cD halos. Evolved stars in these galaxies should inject metals into the hot gas at a rate $\zeta(r)$ proportional to the stellar density (neglecting any radial metallicity gradients for the stars). Since metal injection from stars is equally likely to occur at any point, we assume that the metallicity in a phase of density $\rho$ is increased at a rate $\dot{Z}(r, \rho) = \zeta(r)/\rho$. Thus, the hotter phases become more metal rich than the cooler (and denser) ones. Metals could be produced by type-Ia supernovae or lost from evolved stars with higher metallicity than the gas. The metal-rich ejected gas will have a large kinetic energy (from the supernova explosion or from stellar orbital motions) which will heat the surrounding gas, further increasing the concentration of ejected metals in the hot phases.

The power injected by all the supernovae required to produce an iron abundance enhancement $\Delta Z$ in the cooling flow region is $L_{SN} \sim (\Delta Z \dot{M}/M_{Fe})E$, where $E \sim 10^{51}$ erg and $M_{Fe} \gtrsim 0.3 M_\odot$ are the kinetic energy and iron mass produced by a type Ia supernova (e.g., Woosley & Weaver 1986). The total luminosity of the hot gas in the cooling flow region is $L \sim (5T/2\mu)\dot{M}$, where $T$ is the temperature of the gas, and $\mu = 10^{-24}$ g is its mean molecular weight. Their ratio, $L_{SN}/L \lesssim 0.1(\Delta Z/0.5 Z_\odot)(T/3 \text{ keV})^{-1}$, is small for the Centaurus cluster (for which the reference parameters were chosen), and probably even smaller for richer, hotter clusters, so that the contribution of supernovae to the total energy budget is negligible. Evolved stars will inject gas at their virial velocity, which is comparable to the gas thermal motions. In order for the cooling flow model to be meaningful, the mass of gas injected has to be a small fraction of the total gas mass present (requiring a stellar metallicity much higher than the maximum metallicity of the gas), so that the heat budget is again not substantially altered. Thus, we assume in what follows that only metals are injected, but the mass and energy budget of the cooling flow are unaffected by this process.

If we identify a phase by its "initial" density $\rho_0$ at a fiducial radius $r_0$, its metallicity at radius $r$ is

$$Z(r, \rho_0) = Z_\infty + \int_r^\infty \frac{\zeta(r')\, dr'}{\rho(r', \rho_0)\, u(r')}\,, \qquad (1)$$

where $\rho(r, \rho_0)$ is the density at radius $r$ of the phase of initial density $\rho_0$, and $u(r)$ is the inflow velocity of the gas, assumed to be the same for all phases. $Z_\infty$ is the metallicity at a large distance from the central galaxy, which we assume was laid down before the cluster gas relaxed to its present steady state. In principle, the integral should have an upper cutoff at a radius comparable to the cooling radius, $r_{cool}$ (the radius at which the cooling time equals the age of the cluster), beyond which the steady-state assumption does not hold. Since most of the metals are injected at smaller radii, we will ignore this upper cutoff.

If the total enclosed mass, the gas density profile, and the gas cooling function in the range of interest for the cooling flow are well approximated by power laws, $M(r) \propto r^{\lambda_M}$, $\rho \propto r^{-\lambda_\rho}$, and $\Lambda(T) = \Lambda_0 T^\alpha$, then the cooling flow is self-similar (Nulsen 1986; Waxman & Miralda-Escudé 1995), i.e., the mass inflow rate at radius $r$ of gas in phases with density larger than $\rho$ can be written as $\psi(r, \rho) = \dot{M}(r) I[\rho/\bar{\rho}(r)]$, where $\bar{\rho}(r)$ is the average gas density at radius $r$. The total mass inflow rate is also a power of radius, $\dot{M} \propto r^\delta$, with $\delta = 3 - 2\lambda_\rho - (1-\alpha)(\lambda_M - 1)$. The cumulative density distribution at any given radius is

$$I(\xi) = \begin{cases} [1 - (\xi_{min}/\xi)^{2-\alpha}]^\nu, & \text{if } \xi > \xi_{min}; \\ 0, & \text{otherwise}; \end{cases} \qquad (2)$$

with $\nu = 5\delta/(2-\alpha)/[2\lambda_\rho + 3(\lambda_M - 1)]$.

The density evolution of a given phase is given implicitly by

$$\left[1 - \left(\frac{\rho_{min}(r)}{\rho(r, \rho_0)}\right)^{2-\alpha}\right]^{\nu/\delta} r = \left[1 - \left(\frac{\rho_{0,min}}{\rho_0}\right)^{2-\alpha}\right]^{\nu/\delta} r_0 \equiv r_{dep}(r, \rho)\,, \qquad (3)$$

where $r_{dep}(r, \rho)$ is the deposition radius of the phase whose density is $\rho$ at radius $r$, and $\rho_{min}(r)$ and $\rho_{0,min}$ are the densities of the hottest phase at radius $r$ and at the fiducial radius, respectively. Thus, for this case eq. (1) can be rewritten as

$$Z(r, \rho) = Z_\infty + \frac{1}{\xi_{min}} \int_r^\infty \frac{4\pi r'^2 \, dr' \, \zeta(r')}{\dot{M}(r')} \left[ 1 - \left( \frac{r_{dep}(r, \rho)}{r'} \right)^{\delta/\nu} \right]^{1/(2-\alpha)} . \quad (4)$$

## 3. THE CENTAURUS CLUSTER

Here, we apply the results of § 2 to the Centaurus cluster (A3526), in order to show that a substantial metallicity gradient can in fact be expected in a cooling flow model. We chose this cluster because its radial metallicity distribution has been measured from spatially resolved X-ray spectroscopy by ROSAT (Allen & Fabian 1994) and ASCA (Fukazawa et al. 1994), showing evidence for such a gradient in the cooling region ($r < r_{cool} \sim 6'$, or $60h^{-1}$ kpc, assuming a pure Hubble flow with Hubble constant $H_0 = 100h \, \mathrm{km \, s^{-1} \, Mpc^{-1}}$). However, since the discrepancies between data taken with different instruments are larger than the formal error bars given, it is not clear what errors should be assigned to the data. Thus, we do not attempt a detailed fit, but only a qualitative comparison, and we also omit the error bars from our graphs.

The Centaurus cluster has a bimodal redshift distribution (Lucey, Currie, & Dickens 1986), with two clumps, Cen 30 and Cen 45, each surrounding one of the two brightest elliptical galaxies, NGC 4696 ($cz = 2923 \, \mathrm{km \, s^{-1}}$) and NGC 4709 ($cz = 4519 \, \mathrm{km \, s^{-1}}$). The projected distance between these two galaxies is only $15'$. The X-ray emission of the Centaurus cluster is clearly peaked at the position of the brightest galaxy, NGC 4696, with no evidence for a peak at NGC 4709 (Allen & Fabian 1994).

Figure 1 shows the projected V-luminosity profile of Cen 30. NGC 4696 is approximated as a de Vaucouleurs profile with effective radius $r_e = 1.8'$ and total luminosity $L_V = 8.1 \times 10^{10} h^{-2} L_\odot$ (Schombert 1987). The total V-luminosities of all other galaxies were taken from the data of Dickens, Currie, & Lucey (1986), using their conversions among wavelength bands and assuming de Vaucouleurs profiles for all galaxies.[1] The luminosity distribution in and around the cooling flow region is strongly dominated by the central galaxy, even in projection, making it the likely source of any metallicity excess in this area. We take the metal injection function to be $\zeta(r) = \eta \ell(r)$, where $\eta$ is an adjustable parameter (metal injection per unit time per unit stellar luminosity), and $\ell(r)$ is the V-luminosity per unit volume of the central galaxy, taken as a de Vaucouleurs profile with the parameters given above.

Figure 2 shows the data from the ASCA Gas Imaging Spectrometer (GIS) and Solid-State Imaging Spectrometer (SIS; Fukazawa et al. 1994) and those from the ROSAT Position-Sensitive Proportional Counter (PSPC; Allen & Fabian 1994), together with results from different models. In all of the latter, we assume a power-law gas density profile with $\lambda_\rho = 1.25$, which gives a good approximation to the ROSAT X-ray surface brightness profile (Allen & Fabian 1994). The total mass distribution is not well known, since no lensing observations exist and the temperature profile is fairly uncertain (compare the results of Allen & Fabian 1994 with those of Fukazawa et al. 1994). Thus, we first consider two cooling-flow models which approximately span the range of observationally allowed temperature profiles. In the first, the mass profile, with $\lambda_M = 1 + (3 - 2\lambda_\rho)/(1 - \alpha)$, is chosen to make $\dot{M}$ = constant, corresponding to a single-phase flow. In the second, we use an isothermal mass profile ($\lambda_M = 1$), which gives a multiphase flow with phase-dependent metallicity. Since our model assumes a steady-state cooling flow, which does not apply for $r > r_{cool}$, the metallicity differences among phases shown should be regarded more as indications of the size of the expected effect than as precise quantitative predictions. With the present measurement uncertainties, the disagreement among the data is somewhat larger than the expected iron abundance difference between different phases. This is partly because of the steep surface brightness profile of this cluster, which requires a rather narrow density distribution.

---

[1] This is not accurate in many cases, especially for spiral galaxies, but tends to overestimate the true luminosity of the galaxies. In any case, this correction is typically $< 20\%$, and always $< 41\%$.

In both models, the mass distribution is normalized by the emission-weighted temperature at the cooling radius determined (with low precision) by ASCA, $T_\epsilon(r_{cool}) \sim 3$ keV. Above $\sim 2$ keV, bremsstrahlung is the most important cooling mechanism, and the cooling function $\Lambda(T)$ can be approximated by a power law. Below this temperature, $\Lambda(T)$ increases dramatically due to line emission (e.g., Gehrels & Williams 1993). However, even if $\Lambda(T)$ is approximated by a power law, the cooling time decreases substantially ($\propto T^{2-\alpha}$) as the temperature decreases. Thus, the cooler phases cool quickly and nearly isobarically until they drop out of the flow, and their total emissivity is determined only by the mass deposition rate. Since the mass and volume of the cooling flow are dominated by the hotter phases, inferred properties such as $\dot{M}(r)$ will be unaffected by a change in the cooling function at lower temperatures, as long as the hotter phases remain above $\sim 2$ keV.[2] The metallicity of each phase at deposition is determined by its evolution at high temperatures rather than during the quick cooling through the low-temperature tail of the distribution, so it should be similarly unaffected by this change. Thus, a power-law cooling function should be appropriate for our purpose, namely, to calculate the metallicity as a function of radius, even though the high-density tail of the resulting density distribution is *not* correct. For these low temperatures, the cooling is practically isobaric and the emitted spectrum is independent of the structure of the cooling flow (see Johnstone et al. 1992).

The shape of the metallicity profile is also quite insensitive to the value of $\alpha$. For example, for $\dot{M}$ = constant, it is determined solely by $\zeta(r)$, and for the hottest phase in an isothermal mass profile, it depends on $\zeta(r)$ and $\lambda_\rho$. For definiteness, we choose a power-law cooling function with $\Lambda_0 = 7.8 \times 10^{20}$ erg cm$^3$ g$^{-2}$ s$^{-1}$ $K^{-0.5}$ and $\alpha = 0.2$, which approximates the solar-metallicity curve of Gehrels & Williams (1993) near 3 keV.

In addition to these two standard cooling flow models, we consider two "toy" models with no mass deposition, i.e., in which a heat source is balancing the radiative cooling. In the first, we assume that the gas is static and accumulates locally injected iron over the lifetime of the cluster, so that $Z(r) = Z_\infty + [\ell(r)/\rho(r)] \int_0^t \eta(t')dt'$, where $t$ is the age of the cluster in its present, relaxed state, and $\rho$ is the local gas density. In the second, we assume perfect mixing within the cooling radius, due to a cycle of convection of the hot phases and evaporation of the cooled gas, as mentioned in

---

[2] This is not true at small $r$ in the $\dot{M}$ = constant model, and therefore we expect this approximation to somewhat underestimate $\dot{M}$ and therefore the required $\eta$ for this case.

§ 1.

A visual inspection of the figures shows that the cooling flow models reproduce the observational curves somewhat better than the two extreme models with no net cooling. However, intermediate models with no cooling are possible, and even the extreme models may not yet be ruled out, given the uncertainties in the measured metallicities. The iron injection rates required by all the models considered (see the caption of Fig. 2) are in rough agreement with the observed rate of type Ia supernovae in elliptical galaxies ($1 - 5 \times 10^{-13} h^2 L_{B\odot}^{-1}$ yr$^{-1}$; Turatto, Cappellaro, & Benetti 1994), assuming that each supernova injects a substantial fraction of a solar mass of iron (Nomoto, Thielemann, & Yokoi 1984). However, these rates are substantially higher than those inferred from x-ray observations of the hot gas around some elliptical galaxies not in cluster centers (Serlemitsos et al. 1993; Loewenstein et al. 1994). The expected injection rate from evolved stars is less certain, especially because of the uncertain calibration of stellar metallicity indicators (e.g., Worthey 1994). However, the metallicity of NGC 4696 is high compared to other elliptical galaxies (Davies et al. 1987; Carollo, Danziger, & Buson 1993), implying that this process may well be significant. The relative importance of type Ia supernovae and stellar mass loss could be decided by spatially resolved abundance measurements of other metals. In particular, Fukazawa et al. (1994) have reported that O, Si, Ar, and Ca lines are also stronger in the central region, which would argue in favor of stellar mass loss. It should also be pointed out that a radial decrease in the stellar metallicity has been detected in the central part ($\sim 10''$) of NGC 4696 and other elliptical galaxies (Carollo et al. 1993). If this gradient continues out to arcminute scales, it could steepen the expected metallicity gradient in the gas.

## 4. CONCLUSIONS

The main conclusion we have reached in this paper is that the standard cooling flow model, where gas with the initial metallicity of the intracluster medium flows towards the center of the potential well, with the coolest phases continuously dropping out of the flow, is consistent with the metallicity gradient observed in the Centaurus cluster given reasonable rates of ejection of metals from Type Ia supernovae and evolved stars in the central cluster galaxy (NGC 4696). Most of the ejected metals are deposited in the hot phases, and these are the ones that survive for a longer time in the flow. The gas in the center of the flow is enriched owing to the metals that were deposited in the hot phases at larger radius. This model predicts that, at any radius, the metallicity should increase with the temperature of the phase. This predicts changes in the X-ray spectrum of cooling flows, which might be observable in the future.

Alternative models would predict slightly different metallicity profiles. If the gas was static, the metallicity peak in the center should be more pronounced. In a model where a heating source produces a convection zone, where the gas follows a cycle where it is heated in the center, rises to large radius, and sinks again as it cools, the gas would be mixed and all the metals ejected would be present in the X-ray emitting gas, producing a central "plateau" in metallicity and a sharper decline at the edge of the convection zone. The metallicity would also be more uniform among the phases. However, in the absence of a specific model for a possible heating source, this alternative possibility is not sufficiently predictive to be tested from the metallicity profile.

- 7 -The metallicity enrichment of the cooling flow gas depends mainly on the ratio of the stellar mass in the central galaxy, to the gas mass in the cooling flow. This ratio is relatively large in the Centaurus cluster, and the metallicity gradient should be much smaller in clusters with massive cooling flows where the central galaxy is not very luminous. At the same time, the metallicity gradient should depend on the merging history of clusters: it should be stronger in clusters which have grown quiescently by accreting many small clumps over a long period, without disturbing the radial distribution of the central gas, and weaker in clusters where the central gas has recently been stirred by a large merger. Thus, variations of the metallicity gradient among different clusters are expected.

We thank N. Bahcall, M. Carollo, A. Dressler, B. Gibson, M. Hamuy, A. Loeb, M. Loewenstein, R. Mushotzky, H. Rood, M. Strauss, and S. White for useful discussions, and very specially M. Currie for extensive correspondence and for a computer-readable copy of the Dickens et al. (1986) catalog. We also thank an anonymous referee for comments that helped to improve this paper. The authors are financially supported by NSF grant PHY 92-45317 (A. R. & E. W.) and by the Ambrose Monell (A. R.) and W. M. Keck Foundations (J. M. & E. W.). J. M. was also supported in part by NSF grant PHY 94-07194 at the Institute for Theoretical Physics in Santa Barbara.## REFERENCES

## FIGURE CAPTIONS

**Figure 1:** Enclosed V-luminosity as a function of radius of the Cen 30 redshift clump ($1700\,\mathrm{km\,s^{-1}} < cz < 4150\,\mathrm{km\,s^{-1}}$) in the Centaurus cluster, based on the observations of Dickens et al. (1986) and Schombert (1987). The lines correspond to the luminosity of the central galaxy, NGC 4696 (solid), the early-type galaxies (dashed), and all galaxies (dot-dashed).

**Figure 2:** Iron abundance as a function of radius in the region inside and surrounding NGC 4696. The points correspond to observations with three different instruments, the ROSAT PSPC (Allen & Fabian 1994), and the ASCA GIS and SIS (Fukazawa et al. 1994). The dashed curve is the prediction of the single-phase cooling flow model. The solid curves enclose the range of abundances in different phases in the multiphase cooling flow for an isothermal mass profile, $M(r) \propto r$. The dot-dashed curve is the iron abundance accumulated during the age of the cluster by a static gas. For all three cases, the parameters $Z_\infty$ and $\eta = \eta_{-13} \times 10^{-13} M_\odot L_\odot^{-1}\,\mathrm{yr}^{-1}$ were chosen to fit the SIS data, excluding the innermost point (which may be affected by ASCA's finite spatial resolution; Tanaka, Inoue, & Holt 1994). For the single-phase and $M \propto r$ cooling flow models, $\eta_{-13} = 7.4$ and 3.2, respectively, and for the no-flow model, $H_0 \int \eta_{-13} dt = 1.3$. The dotted line corresponds to perfect mixing of the last model's iron mass within the cooling region.

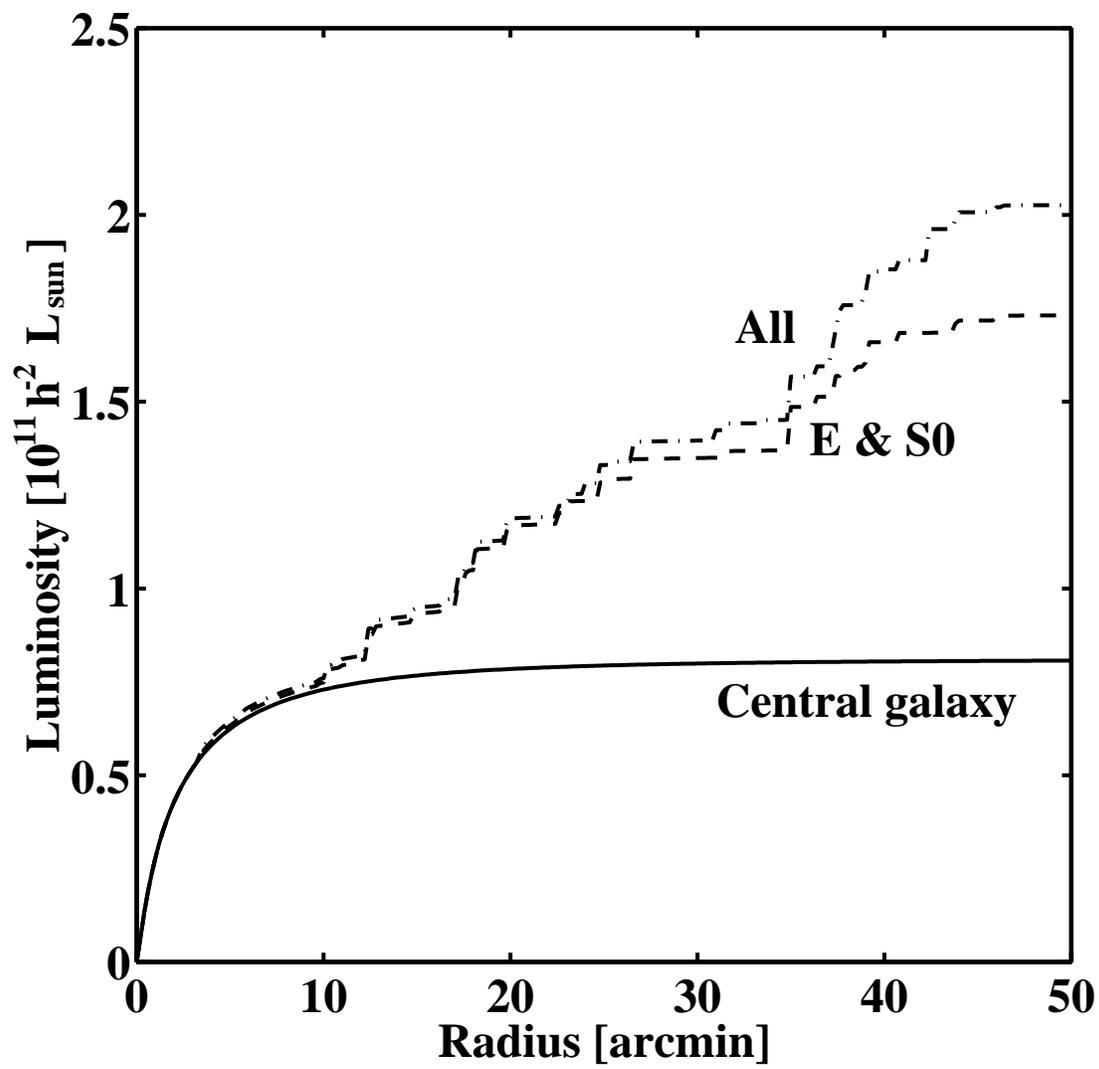

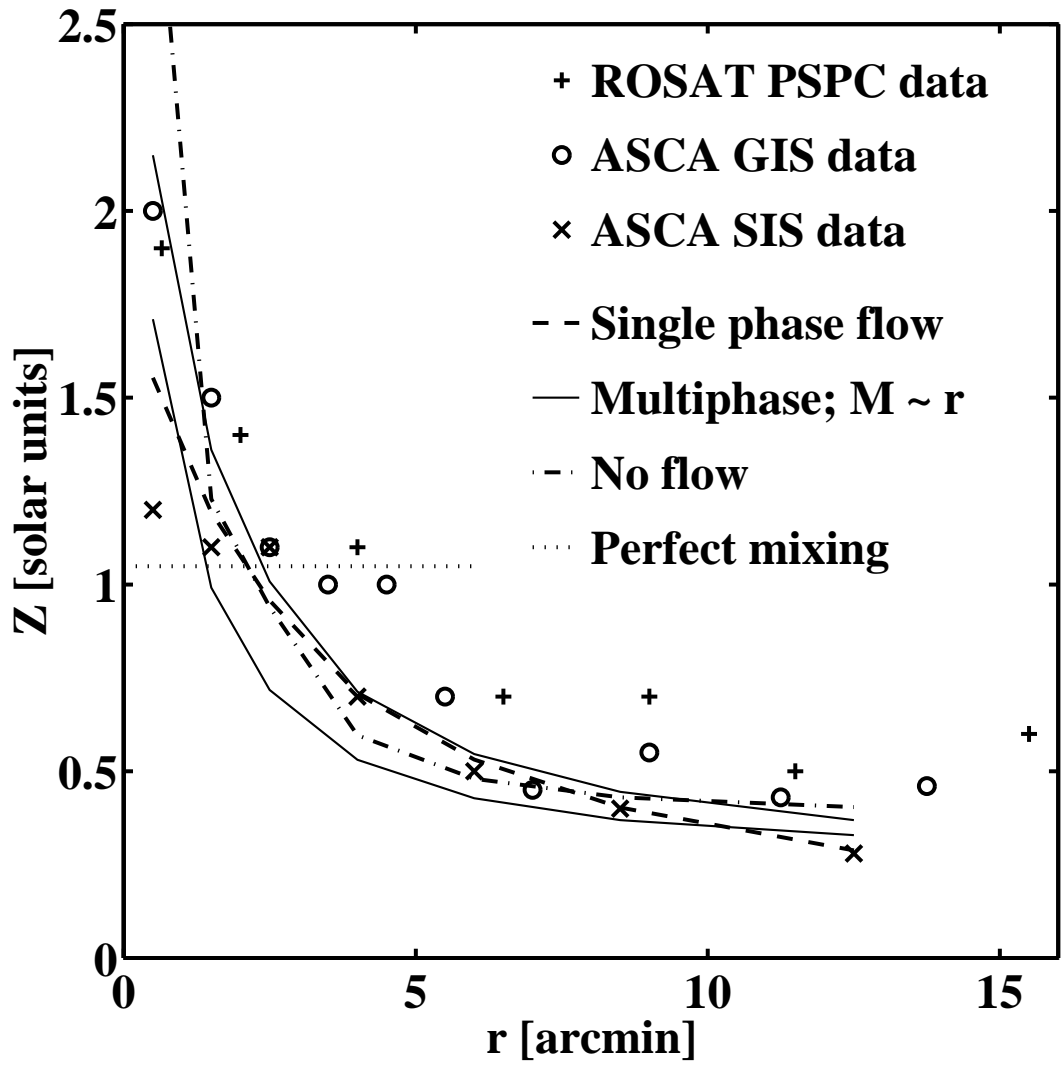